\documentclass[lettersize,journal]{IEEEtran}
\usepackage{booktabs}
\usepackage{longtable}
\usepackage{enumitem}
\usepackage{amsmath,amsfonts}
\usepackage{amsmath}
\usepackage{algorithmic}
\usepackage{algorithm}
\usepackage{array}
\usepackage[caption=false,font=normalsize,labelfont=sf,textfont=sf]{subfig}
\usepackage{textcomp}
\usepackage{url}
\usepackage{verbatim}
\usepackage{color,graphicx}
\usepackage{cite}
\usepackage{xcolor}
\usepackage{dblfloatfix} 
\usepackage{amssymb}
\usepackage{balance}
\usepackage{float}
\usepackage[para,online,flushleft]{threeparttable}
\usepackage{placeins}    
\usepackage{tikz}
\usetikzlibrary{shapes.geometric, arrows}
\newcommand{\probP}{\text{I\kern-0.15em P}}

\begin{document}

\title{\emph{PUFBind}: PUF-Enabled Lightweight 
Program Binary Authentication for FPGA-based Embedded Systems}

\author{Sneha Swaroopa,~\IEEEmembership{Graduate Student Member,~IEEE},
Venkata Sreekanth Balijabudda,~\IEEEmembership{Graduate Student Member,~IEEE}, 
Rajat Subhra Chakraborty,~\IEEEmembership{Senior Member,~IEEE},
and Indrajit Chakrabarti,~\IEEEmembership{Member,~IEEE}
\thanks{S. Swaroopa, V. S. Balijabudda and R. S. Chakraborty are with the
Dept. of Computer Science and Engineering, Indian Institute of Technology 
Kharagpur, Kharagpur, West Bengal, India -- 721302. 
Email: swaroopasneha202@kgpian.iitkgp.ac.in, sreekanthbv@kgpian.iitkgp.ac.in, rschakraborty@cse.iitkgp.ac.in. \newline
I. Chakrabarti is with the Dept. of Electronics and Electrical Communication
Engineering, Indian Institute of Technology Kharagpur, 
Kharagpur, West Bengal, India -- 721302. Email: indrajit@ece.iitkgp.ac.in.}}



\maketitle

\begin{abstract}
Field Programmable Gate Array (FPGA)-based embedded systems have become 
mainstream in the last decade, often in security-sensitive applications. 
However, even with an authenticated hardware platform, compromised 
software can severely jeopardize the overall system security, making 
hardware protection insufficient if the software itself is malicious. 
In this paper, we propose a novel low-overhead hardware-software 
co-design solution that utilizes Physical Unclonable Functions (PUFs) 
to ensure the authenticity of program binaries for 
microprocessors/microcontrollers mapped on the FPGA. Our technique 
binds a program binary to a specific target FPGA 
through a PUF signature, performs runtime authentication for the 
program binary, and allows execution of the binary only after successful 
authentication. The proposed scheme is platform-agnostic and 
capable of operating in a ``bare metal'' mode (no system software 
requirement) for maximum flexibility. Our scheme also does not
require any modification of the original hardware design or program 
binary. We demonstrate a successful prototype implementation using 
the open-source \emph{PicoBlaze} microcontroller on \emph{AMD/Xilinx} 
FPGA, comparing its hardware resource footprint and performance with 
other existing solutions of a similar nature.
\end{abstract}


\section{Introduction}
\label{sec:introduction}

Field-Programmable Gate Array (FPGA)-based embedded systems are 
being increasingly deployed in safety-critical domains such as aerospace, 
defense, power grid infrastructure, transportation, and financial 
systems. The proliferation of these systems is driven by their 
flexibility, and the adoption of industry-grade microprocessors 
and microcontrollers based on open-source instruction sets like 
RISC-V~\cite{ref:Renesas_RISC_V}.  Hence, the privacy and security 
of these hardware systems are critically important due to the 
sensitive nature of the data that they handle, and the severe 
potential consequences of security breaches.

Threats to these critical systems can originate from 
\emph{both} untrusted hardware and compromised software. 
The question that we aim to answer through this paper is: 
\textbf{even with \emph{secure hardware}, how can we ensure 
\emph{overall system security} if the software intended to 
run on it is itself malicious?} This concern is particularly 
critical in scenarios where operating system-enabled security 
mechanisms are absent, such as during the pre-boot phase. 
In several other such ``bare metal'' embedded system 
application scenarios, including industrial and automotive 
control systems, malicious program binaries pose serious threat.

To address this gap, in this paper we have proposed and
demonstrated a novel scheme (which we have named ``PUFBind'') 
that extends the utility of Physical Unclonable Functions (PUFs) 
to include program binary authentication. Physical Unclonable 
Functions (PUFs) are lightweight hardware-security
primitives that generate unique device-specific ``electronic 
fingerprints''~\cite{ref:puf_tutorial}. 
Our proposed solution includes an innovative approach -- 
leveraging PUFs to \emph{bind} a given program binary to 
a target (trusted) FPGA-based hardware platform, ensuring 
the following:
\begin{itemize}
\item the program binary cannot execute on any other FPGA, 
including unauthorized or counterfeit ones, and,
\item only a program binary that has been successfully 
authenticated by hardware means, is allowed to
execute on the target FPGA platform.
\end{itemize}
While binding of the program binary is achieved through 
a PUF signature, the authentication is through the 
industry-standard SHA-256 cryptographic hash function. 
The combination of the PUF signature and the SHA-256 
digest places the proposed authentication scheme on sound 
security foundations, linking the program binary and the 
FPGA-platform by cryptographic means. 

\emph{PUFBind} consists of two main steps: 
%
\begin{enumerate}
\item \textbf{Step-1 (Trust establishment of FPGA platform):}
A PUF circuit is mapped on the target FPGA, and its signature noted.
Later, an FPGA-under-test is authenticated by implementing the same
PUF circuit on it, and noting its signature. If the obtained 
PUF signature matches the expected PUF response, the FPGA is declared
to be authentic. Error Correction Codes (ECCs) are employed to 
address reliability issues associated with PUF responses.

\item \textbf{Step-2 (Authentication of program binary on a 
pre-authenticated FPGA platform):} This step consists of several
sub-steps, as described below.
\begin{itemize}
\item \emph{Binding a Given Program Binary to a FPGA:} 
The PUF response from the target FPGA is used to cryptographically 
bind a program binary to the hardware. A SHA-256 digest of a given
program binary is XOR-ed with the PUF signature from the 
trusted target FPGA, to obtain a ``golden signature''. This golden
signature, representing the authentic program binary and its intended
target FPGA, is then piggybacked on the program binary 
file itself. This sub-step is a one-time effort for each 
$<$program binary, FPGA$>$ pair. 
\item \emph{Verification of Program Binary:} Prior to allowing
a given program binary to execute on a soft microprocessor/microcontroller
mapped on a trusted FPFGA, the program binary is authenticated in real time
by hardware. If authentication passes, normal execution ensues.
Otherwise, any further reading of the memory containing 
this program binary is disabled by hardware controllers, which thereby 
forbids any further execution of it.
\item \emph{Trustworthy Program Binary Execution:} A complementary
scenario arises when the program binary is trustworthy,
but the FPGA on which it is intended to run is unauthorized, i.e., not
the intended FPGA to which the program binary was bound. Again, 
a real-time hardware-based authentication mechanism disables 
execution of the program binary. 
\end{itemize}
\end{enumerate}




\begin{figure}[!t]
\centering
\includegraphics[scale=0.57]{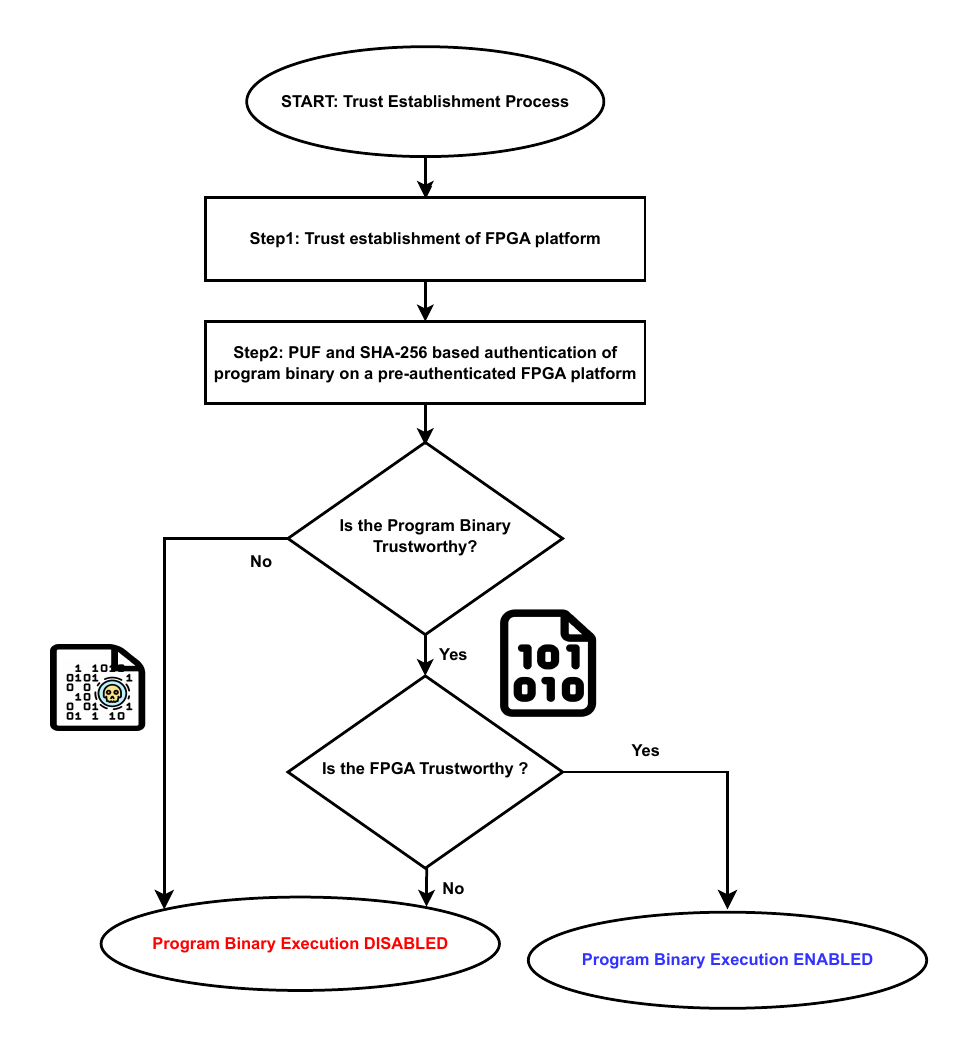}
\caption{The different authentication scenarios and the corresponding outcomes,
as enforced by \emph{PUFBind}.}
\label{fig:PUFBind_scenarios}
\end{figure}
Fig.~\ref{fig:PUFBind_scenarios} capture the different scenarios and 
the corresponding outcomes, as enforced by \emph{PUFBind}. 

\subsection{Salient Features of \emph{PUFBind}}
\label{sec:salient_features}
We have made the following security assumption: 
\textbf{an adversary aiming to maliciously
modify the program binary, does not have access to the FPGA platform(s)
to which the binary is securely bound.} Under this assumption,
\emph{PUFBind} offers the following:
\begin{itemize}
    \item \emph{PUFBind} eliminates the need  for encryption
    followed by run-time decryption of the program binary,
    thus avoiding adverse impact to throughput.
    \item \emph{PUFBind} operates without requiring a complex 
    software stack, in a \emph{bare metal} mode.
    \item \emph{PUFBind} does not place any limit on the 
    program binary size. 
    \item \emph{PUFBind} is completely \textbf{non-invasive} 
    -- it requires no modification to the hardware modules 
    to be mapped on the FPGA, and requires only minimal 
    and trivial modifications to the program binary 
    Hence, there is no need to separately store 
    the authentication signature in non-volatile memory. 
    \item The hardware platform (FPGA) authentication is a one-time 
    operation, carried out before the program binary begins
    execution on the FPGA. 
    It has no adverse impact on the throughput of the 
    normal functional units. 
    \item \emph{PUFBind} seamlessly supports authorized
    updates to the program binary and porting the given program 
    binary to another target FPGA platform - only the 
    initial one-time binding process needs to be re-executed.
    \item \emph{PUFBind} has a very small 
    resource footprint (please refer to Table~\ref{table:ta}). 
    Since the authentication unit can be powered-down after 
    successful authentication, it is also power-efficient.
    \item Lastly, \emph{PUFBind} ensures the complete 
    satisfaction of the CIA triad of security —
    \textbf{\emph{Confidentiality}, \emph{Integrity}}, and 
    \textbf{\emph{Availability}}. 
    \emph{Confidentiality} is achieved indirectly 
    by binding the program binary to authorized hardware 
    using PUF-based authentication, ensuring that only secure, 
    trustworthy hardware can access (read from memory) 
    and execute the program binary, 
    while preventing access or execution on untrusted devices. 
    \emph{Integrity} is maintained through cryptographic hashing, 
    which detects even a single bit modification to the binary 
    and prevents execution if tampering is detected. Finally, \emph{Availability} is ensured by enabling execution 
    immediately after successful pre-execution authentication, 
    guaranteeing the program runs as intended without additional 
    delays.
\end{itemize}

\textbf{To the best of our knowledge, this 
is the first work to leverage PUFs for 
authenticating program binaries before 
execution~\cite{ref:security_in_FPGA_based_systems}}

\subsection{Organization of the Paper}
The rest of the paper is organized as follows. In 
Section~\ref{sec:background}, we provide necessary 
background of program binary authentication, PUFs, 
PUF-based secure system development,
and our contribution with respect to state-of-the-art.
In  Section~\ref{sec:proposed_sceheme}, we describe the 
PUF design and the proposed scheme in detail, and
provide a security proof. In
Section~\ref{sec:results}, we present hardware resource
estimates for a prototype implementation of the proposed scheme 
and compare the results with those of existing 
alternatives for program binary authentication. 
We conclude in Section~\ref{sec:conclusions}, 
with future directions of research.

\section{Background}
\label{sec:background}

\subsection{Program Binary Integrity and Its Impact}
At current state-of-the-art, it is relatively straightforward 
to disassemble (reverse-engineer) a program binary using
powerful (often free-of-cost) disassemblers~\cite{ref:disassembler_1}, 
make malicious modifications by adding or modifying instructions 
\cite{ref:disassembler_2}, and then assemble it back again to 
another valid program binary. Alternatively, well-directed faults 
can be injected to make malicious modifications to program binaries
\cite{ref:variable_length,ref:fault_injection}. 
If the modification of a program binary is undetected, and a 
compromised program binary is deployed in-field, it can 
cause severe functionality losses or 
disruptions~\cite{ref:secure_firmware_update}. For instance, 
the \emph{Stuxnet} malware~\cite{ref:stuxnet} caused 
physical damage to Iran’s nuclear facilities, while the 
\emph{SolarWinds} cyber-attack~\cite{ref:Solarwinds} led 
to the surreptitious leakage of large volumes of sensitive 
information. More recently, compromised communication 
devices (pagers) in possession of members of an 
internationally-recognized
terror-outfit, exploded after receiving deceptive 
specially-encoded messages, which acted as malicious triggers
\cite{ref:hezbollah_explosions}. It can be envisaged
that the modified firmware in these compromised devices, on
recognizing the trigger messages, executed unintended 
and catastrophic commands, resulting in physical damage 
and multiple explosions of the devices.  

\subsection{Existing Schemes for Program Binary Authentication}

\subsubsection{Remote System Monitoring and Attestation}
Faced by the abovementioned challenges, protocols for 
secure software (including firmware) updates and remote 
device and firmware attestation have become major areas of 
research~\cite{ref:secure_firmware_update}. These 
have led to the development and standardization of advanced 
and complex protocols for system state monitoring and attestation, 
encompassing both hardware and deployed firmware/software, 
e.g., through \emph{Security Protocol and Data Model} 
(SPDM)-based schemes~\cite{ref:SPDM_formal_analysis}.

\subsubsection{Hardware-assisted Software and Platform Authentication}
Another popular solution option is to adopt an integrated 
solution that entwines the security of software \emph{and} 
the hardware platform on which it will execute, e.g., 
the widely used hardware-enabled \emph{Trusted Platform 
Module} (TPM)~\cite{ref:TPM_book,ref:TPM_intel},
\emph{Secure Enclave}~\cite{ref:secure_enclave_apple}
for Apple platforms, and the \emph{TrustZone}~\cite{ref:trustzone_arm} technology 
from ARM. 
Among the existing solutions, these security schemes 
are the closest to our proposed PUF-based program binary 
authentication scheme. However, they are far more complex,
often platform-specific, incapable of operating in a bare-metal
mode, require static \emph{in situ} storage of keys, and are 
often not suitable for resource-constrained devices
(except \emph{TrustZone}).
Table~\ref{tab:comparison} summarizes
the differences between \emph{PUFBind} and these schemes.
Other sophisticated but complex commercially available 
schemes for firmware authentication
include~\cite{ref:ti_firmware}. 

\subsection{Confidential Computing Solutions}
\emph{Confidential Computing} techniques~\cite{ref:AEGIS_processor}, 
such as the \emph{Intel Trust Domain Extensions} 
(Intel TDX)~\cite{ref:intel_tdx}, enable the creation of 
tamper-evident and private execution environments 
via hardware-based isolation in computer systems with advanced
multi-core processors. These solutions leverage 
instruction-set extensions and specialized hardware accelerators
to securely deploy virtual machine instances even 
in untrusted environments, including untrusted operating 
systems and hypervisors. 
Again, our proposed approach differs fundamentally from 
Confidential Computing schemes in many regards, including
avoiding the need for instruction-set extension or
vendor-specific hardware. Also, our proposed technique 
is highly adaptable across a diverse range of systems. 
\begin{table*}[!t]
\centering
\caption{Comparison of Existing Hardware-assisted Schemes and Proposed Scheme for Program Binary Authentication}
\scalebox{0.95}{%
\begin{tabular}{@{}p{2cm}p{3cm}p{3.5cm}p{3cm}p{3cm}@{}}
\toprule
\textbf{Aspect} & \textbf{Secure Enclave ~\cite{ref:secure_enclave_apple} } & \textbf{TPM ~\cite{ref:TPM_book,ref:TPM_intel}} & \textbf{TrustZone ~\cite{ref:trustzone_arm} } & \textbf{\emph{PUFBind} (proposed)} \\ \midrule

\textbf{Platform Dependency} & Platform-specific, tightly integrated with Apple’s hardware & Platform-agnostic but requires TPM hardware support, common in Windows-based systems & Tightly coupled to Arm architecture, requires specific Arm processors with TrustZone architectural extensions & Platform-agnostic, adaptable to various hardware and operating systems without vendor-specific dependencies. \\ \midrule

\textbf{Vulnerabilities} & Potentially vulnerable due to shared memory keys between Secure Enclave and Secure Neural Engine. If memory protection isn’t properly enforced, 
shared memory space may be exploited. & Prone to attacks like buffer overflows (CVE-2023-1017 and CVE-2023-1018), risking cryptographic keys and data. & No separation of normal and secure world resources. Vulnerable to malicious binaries in the secure domain due to limited hardware-based isolation and resource sharing. & Eliminates static storage of keys, dynamically generates PUF responses using device-specific physical variations, reducing risk. \\ \midrule

\textbf{Resource Requirements} & Requires dedicated hardware resources integrated into the system’s chipset, adding to cost and power. & TPMs require dedicated hardware, either as a discrete chip or integrated within the CPU. This dedicated TPM hardware, increases cost and power consumption, especially for smaller devices. & Leverages Arm TrustZone extensions within the processor, avoiding extra hardware but sharing resources, limiting advanced isolation. & Leverages intrinsic hardware characteristics, no need for additional hardware modules, reducing cost and complexity. \\ \midrule

\textbf{Performance} & Can experience performance degradation during cryptographic operations, including key generation and memory encryption/decryption tasks. The need for secure memory and cryptographic operations may affect performance during execution. & Hardware TPMs can throttle under high loads, such as intensive key generation or non-volatile memory operations. & Limited to two operational domains (secure and normal), reducing overall performance for systems requiring multi-domain isolation or complex secure operations. & \textbf{No Runtime Overhead:} Authentication occurs only once before the program binary begins execution. Once authentication is completed, there’s no need for runtime decryption, resulting in no performance bottlenecks. This leads to optimal throughput in FPGA-based systems without the continuous overhead of encryption/decryption. \\ \midrule

\textbf{Scalability} & Limited scalability across platforms due to platform-specific implementation. & Limited scalibility in resource-constrained environments, such as
IoT devices, due to cost and power
overheads. Limited usability as
TPM is typically tied to specific
devices and platforms which has
TPM hardware support. & Not inherently scalable due to dependence on Arm architecture and the inability to handle more than two payloads or cores simultaneously. & Highly scalable across various devices and platforms. Lightweight and low-cost, ideal for IoT and embedded systems, as it requires no specific vendor, architecture, processor, and operating system dependencies. \\ \midrule

\textbf{Flexibility} & No flexibility due to tight integration with Apple’s ecosystem. & Limited flexibility, mainly for devices that support TPM modules. & Limited flexibility, restricted to specific Arm-based processors with TrustZone extensions. & Very flexible, capable of working across various devices and systems without requiring proprietary hardware or software. \\ \midrule

\textbf{Key Management} & The Secure Enclave generates and manages keys in a secure environment, with no key export capabilities. & TPM generates and manages keys for platform-specific tasks, can export keys for external use under certain conditions. & Uses the secure world for key management, relying on trusted software, but lacks inherent mechanisms to prevent malicious binary execution if the secure world is compromised. & PUF signatures are derived from unique physical characteristics of the hardware, eliminating static key storage, with no need for explicit key management. \\ \midrule

\textbf{Use Cases} & Secure boot, encrypted storage, and authentication for Apple devices, ensuring confidentiality through encryption. & Secure boot, key exchange, disk encryption, and authentication for enterprise systems, focused on confidentiality. & Secure execution environment for trusted applications with separation of normal and secure worlds, but limited for multi-domain operations, ensuring confidentiality. & Ensures confidentiality through PUF-based binding to authorized hardware, while integrity and authenticity 
are guaranteed by cryptographic hashing. \\ \bottomrule

\end{tabular}}
\label{tab:comparison}
\end{table*}
\subsection{PUF and PUF-based Secure System Development}
PUF~\cite{ref:puf_Gassend2002,ref:puf_tutorial} has 
emerged as a popular low-overhead hardware security primitive, 
serving as an alternative to 
traditional cryptographic primitives. PUFs allow secure 
systems to be built with no (or minimal) storage of secure 
keys \emph{in situ}. The operation of PUFs relies on the 
random and unpredictable impact of manufacturing process 
variation on the nanoscale transistors and interconnects 
in integrated circuits, to generate unique identifiers or 
``electronic fingerprints'' for ICs. PUF technology has matured to a 
point where silicon-ready PUF intellectual property (IP) 
modules, which can be directly integrated to be part of 
any larger design, are available from leading 
vendors~\cite{ref:puf_synopsys,ref:neopuf,ref:pufcc}.
PUFs have been used in a plethora of security applications,
as described below~\cite{ref:synopsys_puf,ref:pufcc}: \par
\begin{figure*}[!t]
\centering
\includegraphics[scale=0.60]{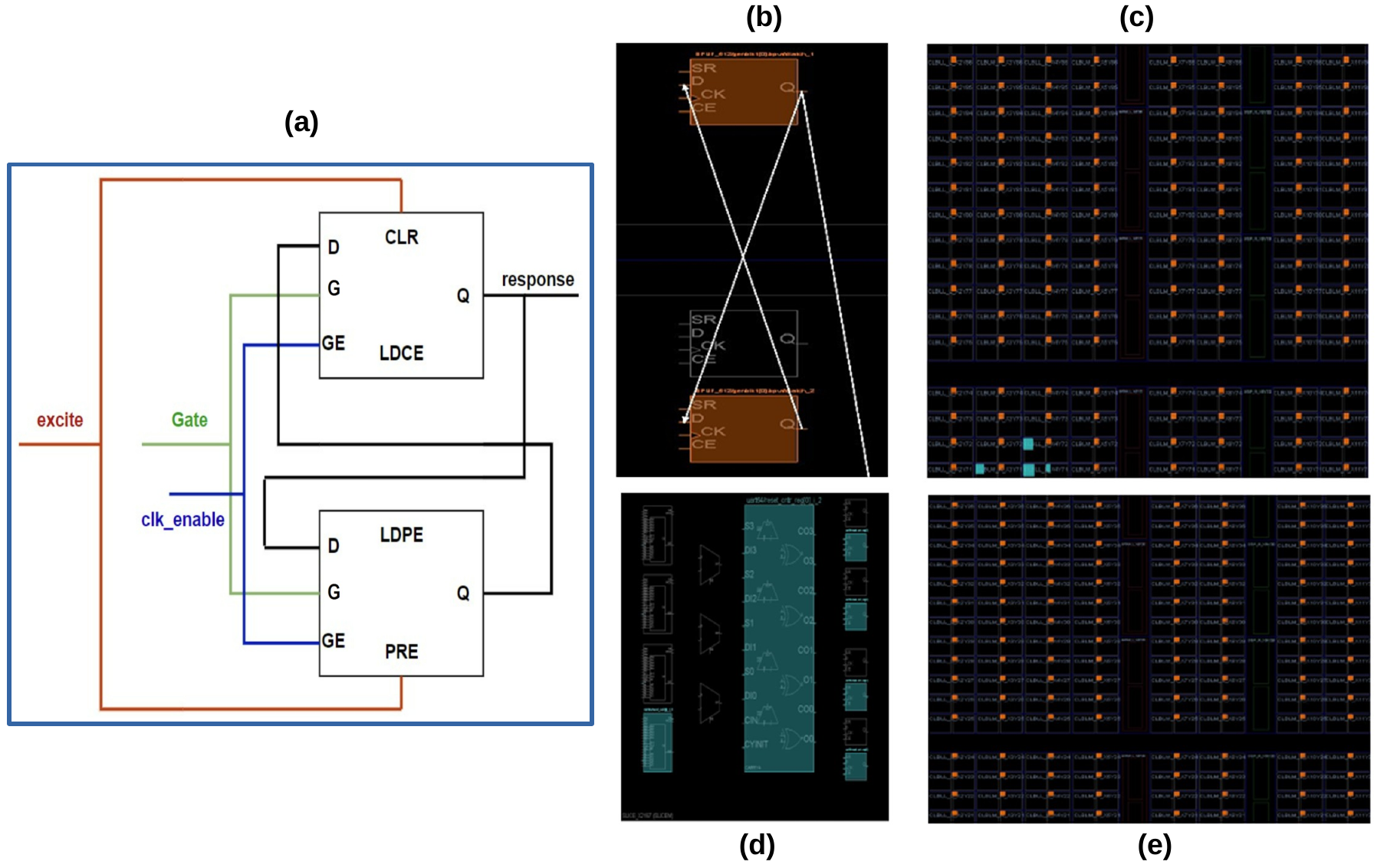}
\caption{Details of Butterfly PUF (BPUF) implementation on \emph{Xilinx
Artix-7} FPGA: (a) circuit schematic of a single BPUF cell showing 
cross-coupled latches (implemented with Xilinx-specific hardware
modules \texttt{LDPE} and \texttt{LDCE}); (b) physical view of
implementation of a single BPUF cell on \emph{Xilinx/AMD Artix-7} FPGA; 
(c) physical view of symmetric placement of first 64 BPUF cells (out of 128) 
on the FPGA fabric; (d) physical view of control circuitry for the BPUF 
circuit on the FPGA fabric; (e) physical view of symmetric placement 
of the second 64 BPUF cells (out of 128) on the FPGA fabric.}
\label{fig:butterfly_PUF}
\end{figure*}
\textbf{1. PUF-based secret key protection (``Key vault")
\cite{ref:synopsys_puf_vault}:} An input user key is
cryptographically ``wrapped'' using a PUF response, and the wrapped key 
is then used to encrypt sensitive information. Later,
the same procedure is used to generate the wrapped
key that can decrypt the encrypted information. \newline
\textbf{2. PUF-based hardware IP locking~\cite{ref:synopsys_firm_IP}:}
The ``firm'' (gate-level) IP is encrypted using a PUF
generated key, and can only be unlocked in the correct
hardware instance (IC), thus providing protection against
reverse-engineering. Additionally, the technique provides protection
against cloning and overbuilding, as the firm IP module
locked with a given PUF key cannot be unlocked by a different
PUF key sourced from a different (cloned or overbuilt) IC. \newline
\textbf{3. Cryptographic co-processors including PUF
\cite{ref:synopsys_puf_software,ref:pufcc}:} These
are comprehensive solutions that include PUF, True 
Random Number Generator, a complex hardware 
cryptographic engine capable of performing key 
derivation, key wrapping, symmetric key encryption/decryption, 
cryptographic hash functions, message authentication code, 
etc.

It is to be noted that in many recently proposed 
applications PUFs do not completely replace traditional cryptographic 
operations, but are used in conjunction 
with them (e.g. see~\cite{ref:puf_IoT_protocol}. We also adopt a 
similar approach, where we use PUFs in conjunction with cryptographic 
hash functions. 

Next, we describe the \emph{PUFBind} scheme in detail.

\section{\emph{PUFBind}: Details of the Scheme}
\subsection{Butterfly PUF: FPGA Implementation and 
Reasons for Choosing It}
The \emph{Butterfly PUF} (BPUF) is a relatively simple and widely-used
PUF circuit, originally proposed in~\cite{ref:butterfly_PUF}.
Fig.~\ref{fig:butterfly_PUF}(a) shows the circuit schematic of a
single PUF cell, as implemented using hardware primitives available
on our FPGA platform of choice: the \emph{Xilinx (AMD) 
Artix-7}. Note that while a SRAM cell consists of cross-coupled 
inverters, the BPUF chose to use cross-coupled latches, 
to bypass the difficulty of implementing combinational loops
on a FPGA.

The implementation of each BPUF involves arranging two D-latches 
(upper one with asynchronous clear, and lower one with asynchronous preset) 
in a cross-coupled manner. The \texttt{excite} signal is connected 
to the clear of the upper and preset of the lower latch. The 
\texttt{Gate} (G) input of each latch is connected to the 
\texttt{gate} signal, and the \texttt{Gate enable} (GE) input 
is connected to the \texttt{clk\_enable} signal. These two 
signals are enabled periodically. In the beginning, when the 
\texttt{excite} signal is made high for some clock cycles, the 
\texttt{Q} output of the upper latch is forced to become `0', 
and that of the lower latch is forced to become `1'. Before making the 
\texttt{excite} signal low again, the \texttt{gate} signal 
is made high, so that the \texttt{D} to \texttt{Q} transfer 
is possible in Latch if the \texttt{GE} signal 
is  activated, and after that, the \texttt{clk\_enable} signal 
is also made high. Now, when the \texttt{excite} is made low, 
an unstable state occurs when the outputs of the latches hold 
opposite values, but since the \texttt{D} input of the 
upper latch is connected to the \texttt{Q} output of the lower
latch, and D of lower is connected to Q of upper, so at 
the same time, outputs tend to hold same values. Now, when the 
\emph{gate} is made low, the D to Q transfer in the last negative 
edge is taken into account.
The exact value at which the output \texttt{response} settles is
determined by small differences in the drive-strengths of the 
latches, arising out of unpredictable and uncontrollable 
silicon-level process-variation effects. The process variation 
also varies spatially across the FPGA fabric, resulting in a 
characteristic bit-patten for an array of BPUF cells, which act 
as the signature for the FPGA. To be practically useful, typically 
we require a BPUF signature to be
length at least 128 bits. In addition, to eliminate design bias, it is 
important that the physical implementation of a BPUF cell should 
have symmetric routing of the interconnections, and the BPUF
cells of the array should be placed symmetrically across the
FPGA design fabric.

\begin{figure*}[!t]
\centering
\includegraphics[scale=0.45]{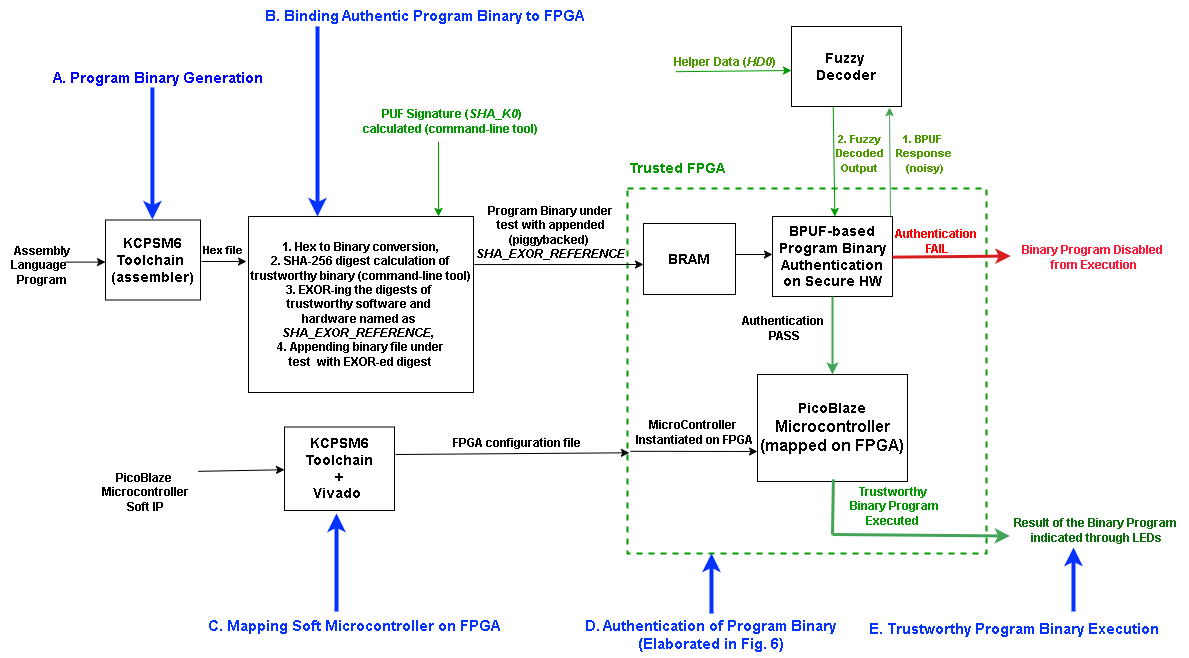}
\caption{Overall architecture of PUF-based program binary authentication
through \emph{PUFBind}, for a prototype implementation on 
\emph{Xilinx (AMD)} FPGA.}
\label{fig:Overall_Architecture_PUFBind}
\end{figure*}
We chose the BPUF in our work 
because it is relatively easy to implement on any FPGA platform as it 
requires only latches,  
has excellent performance metrics (e.g. uniqueness), and incurs 
low hardware overhead (the other
similar option that we considered is~\cite{ref:anderson_puf}).
The \emph{SRAM PUF}~\cite{ref:puf_synopsys} and the
\emph{NeoPUF}~\cite{ref:neopuf} may be overall of more 
superior quality that the Butterfly PUF. However, the SRAM PUF is 
difficult to implement on a FPGA, since the automatic bit-clearing 
mechanism on power-up for all on-chip SRAM of the
FPGA needs to be bypassed~\cite{ref:sram_puf_fpga}. The NeoPUF
is more suitable for a custom IC implementation,
and is only available (to the best of our knowledge) 
as a ``hard IP'' targeting FPGAs from a single 
vendor~\cite{ref:neopuf_on_fpga}, 
which is not portable across other FPGA platforms. The entropy and
reliability of our PUF response is enhanced by ``Fuzzy Extraction" 
for error correction, using the widely-used ``helper data" algorithm
\cite{ref:puf_helper_data_algo,ref:fuzzy_extractor_theory}.
Note that a Butterfly PUF hardware IP solution is also commercially
available targeting AMD/Xilinx FPGAs~\cite{ref:synopsys_butterfly_puf}.
Fig.~\ref{fig:butterfly_PUF}(b)-(d) show the physical view for 
different components of an 128-bit BPUF circuit implemented on FPGA,
with controlled routing and placement, which was achieved through
appropriate design constraints (details in Section~\ref{sec:results}).

\begin{figure}[!t]
\centering
\includegraphics[scale=0.50]{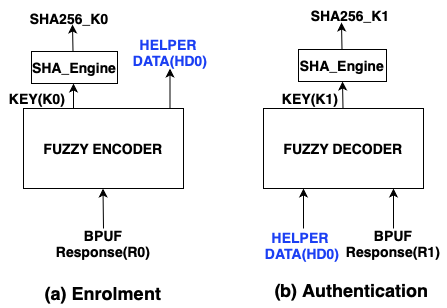}
\caption{PUF-based FPGA platform trust establishment.}
\label{fig:Hardware_Authentication}
\end{figure}

Fuzzy 
Extractor~\cite{ref:fuzzy_extractor_theory,ref:fuzzy_extractor_software} 
is an advanced cryptographic tool used to derive reliable keys 
from noisy sources. 
The implementation employs Hamming distance as the error 
metric to generate the same key with high probability if 
the binary strings' distance is within a specified threshold. 
In our design, we utilized a Python-based 
software implementation of the Fuzzy Extractor, as described 
in~\cite{ref:fuzzy_extractor_software}. The Fuzzy Extractor is 
instantiated with two key parameters: the input size \( n \) and 
the number of error bits \( k \) that can be corrected with high
probability. Specifically, in our case, the input to the 
Fuzzy Extractor is \( n = 128 \), and the error correction 
capability is set to \( k = 13 \), representing the approximate 
number of bits the Fuzzy Extractor can reliably (i.e., with
high probability) correct. It should be noted that \( k \) 
may vary around this value due to the inherent probabilistic 
nature (\emph{fuzziness}) of the system. The 128-bit output 
generated by the BPUF is provided as input to the Fuzzy 
Extractor, along with auxiliary information termed 
``helper data'' to aid in error correction
\cite{ref:puf_helper_data_algo,ref:fuzzy_extractor_theory}, 
to produce a secure output key. This key is then truncated 
to 448 bits and concatenated with 64 padding bits as per the SHA-256
specifications,
resulting in a 512-bit input. The SHA engine processes 
this 512-bit input and computes a 256-bit SHA digest as the 
final output. The overall process is illustrated in 
Fig.~\ref{fig:Hardware_Authentication}.


\subsection{Steps of \emph{PUFBind}}
\label{sec:proposed_sceheme}
As mentioned in Section~\ref{sec:introduction}, 
\emph{PUFBind} consists of two main steps:
\begin{itemize}
\item \textbf{Step-1}: Trust establishment of FPGA 
platform (hardware authentication), 
and,
\item \textbf{Step-2}: PUF-based authentication of program 
binary on a pre-authenticated FPGA platform. It consists
of the following sub-steps 
(as marked in Fig.~\ref{fig:Overall_Architecture_PUFBind}):
\begin{enumerate} [label=\Alph*.]
    \item \emph{Program Binary Generation}
    \item \emph{Binding Authentic Program Binary to FPGA} 
    \item \emph{Mapping Soft Microcontroller on FPGA}
    \item \emph{Authentication of Program Binary}
    \item \emph{Trustworthy Program Binary Execution}
\end{enumerate}
\end{itemize}

In our description below, for ease of understanding, 
we consider a prototype implementation consisting 
of the \emph{Xilinx/AMD Atrix-7} FPGA hosted on an 
\emph{Digilent Nexys-A7} FPGA board, 
the 8-bit \emph{PicoBlaze} open-source microcontroller
made available as the \emph{KCPSM6} soft macro IP 
free-of-cost by Xilinx/AMD~\cite{ref:AMD_picoblaze},
and the \emph{Xilinx/AMD Vivado} FPGA design environment. 
\textbf{However, note that this is a completely generic 
and flexible scheme that can be adapted to any FPGA from 
any vendor on which a BPUF or a similar high-quality 
PUF can be implemented.}  
\begin{figure*}[!t]
\centering
\includegraphics[scale=0.85]{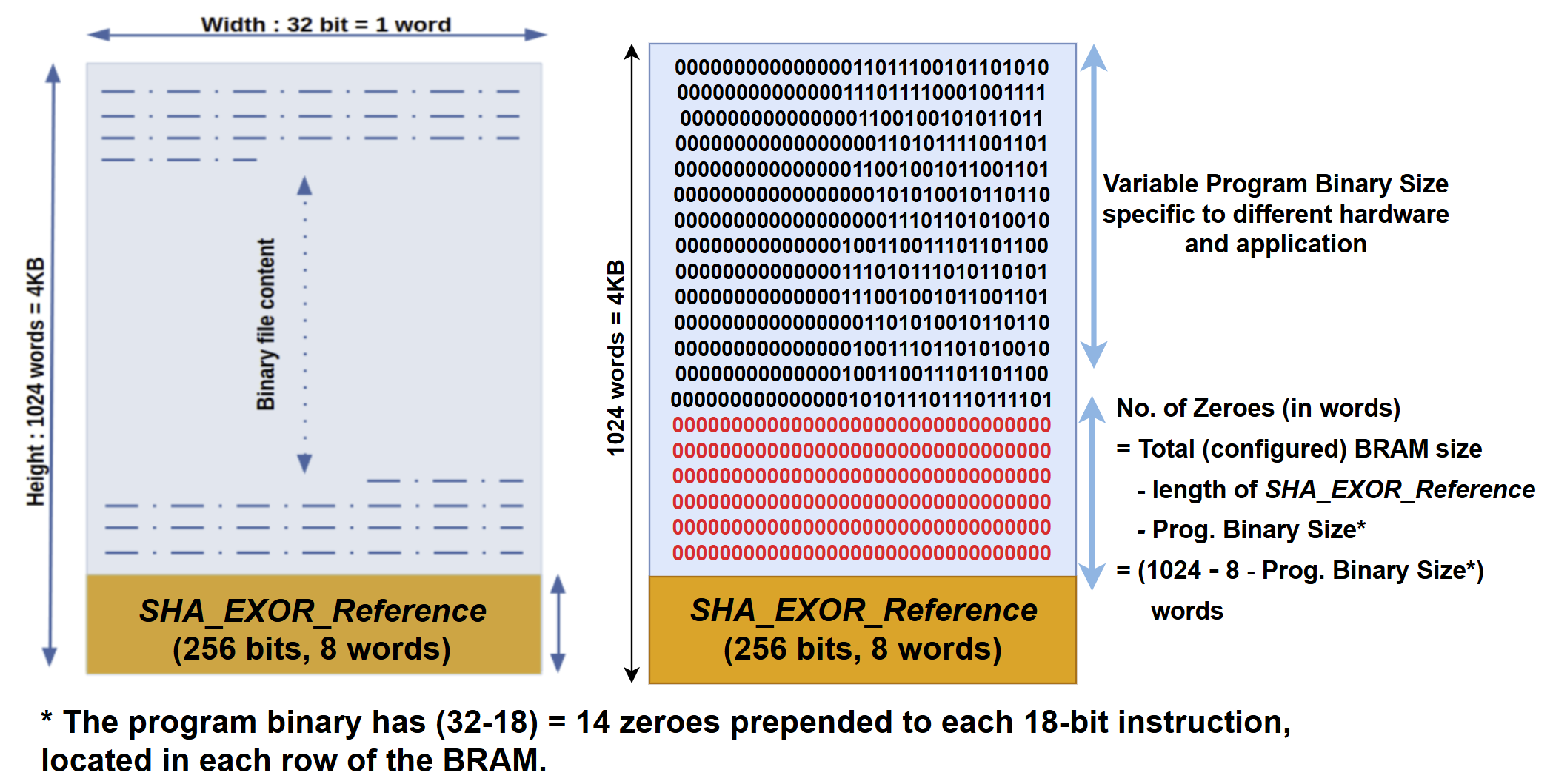}
\caption{Example program binary image and BRAM content, for a 
BRAM with 32-bit words and total size 4 kB.}
\label{fig:Bin_File}
\end{figure*}

The KCPSM6 distribution comes with the software tools 
necessary for complete system development (including an 
assembler to generate program binaries targeting the 
microcontroller mapped on a Xilinx FPGA), and also 
includes an Universal Asynchronous Receiver/Transmitter
(UART) hardware core for easy communication. 

\subsubsection{Step-1: Trust Establishment of FPGA 
Platform (Hardware Authentication)} 
\label{sec:trust-establishment}
 The FPGA platform authentication is achieved using a PUF-based 
 identifier unique to each FPGA chip. This step, referred to as 
 ``FPGA enrolment'' is a one-time process performed by the FPGA 
 manufacturer for each individual FPGA before it is released to the
 supply chain. Fig.~\ref{fig:Hardware_Authentication} shows
 this step. A BPUF array with a pre-determined number of response bits, 
 and pre-determined placement on the FPGA, is mapped on the FPGA, and 
 its response ($R0$) is recorded on a computer through an 
UART interface. Then, the Fuzzy Extractor is applied on $R0$. 
The Fuzzy Extractor operates in two distinct phases: 
the \emph{Encoding} phase and the \emph{Decoding} phase. The 
Encoding phase is responsible for 
generating a high-entropy ``key'' ($K0$), along with auxiliary 
public information termed as ``helper data'' ($HD0$)
\cite{ref:puf_helper_data_algo,ref:fuzzy_extractor_theory}.

While lightweight hardware implementations of the
helper data scheme has been proposed
\cite{ref:puf_helper_data_algo}, a hardware implementation
is not essential, and
we found an easy-to-use available software alternative 
(e.g.~\cite{ref:fuzzy_extractor_software}) suitable
for our scheme.
Finally, a SHA-256 cryptographic digest of $K0$ 
is generated ($SHA256\_K0$), 
by using a standard Linux command-line tool. We used the
command-line SHA-256 digest calculation method in 
Step-1 to corroborate its results with the output 
generated by the hardware SHA-256 engine, thereby
establishing the functional correctness of the SHA-256
hardware engine. The value $SHA256\_K0$ is used as 
the identifier for the FPGA instance.
The tuple $<\!\!SHA256\_K0$, $HD0\!\!>$ 
needs to be passed unchanged from vendor to vendor 
across the IC supply chain, up to the 
final customer, as an incorruptible unique identifier for the FPGA.
To facilitate the FPGA authentication phase 
(described below), the authenticator will be 
provided the BPUF hardware IP. 

A party in possession of the tuple $<\!\!SHA256\_K0, HD0\!\!>$
and a FPGA chip corresponding to it, can perform ``FPGA 
authentication'', by implementing the BPUF on the FPGA, 
collecting its response ($R1$), and in conjunction with 
the helper data $HD0$, generates a key $K1$ by Fuzzy 
Extractor Decoding (again by using a software mean).
If the SHA-256 digest of $K1$, denoted by 
$SHA256\_K1$, matches $SHA256\_K0$, the FPGA 
is authenticated.

\subsubsection{Step-2: PUF-based Authentication of 
Program Binary on a Pre-authenticated FPGA Platform} 
\begin{figure*}[t]
\centering
\includegraphics[scale = 0.16 ]{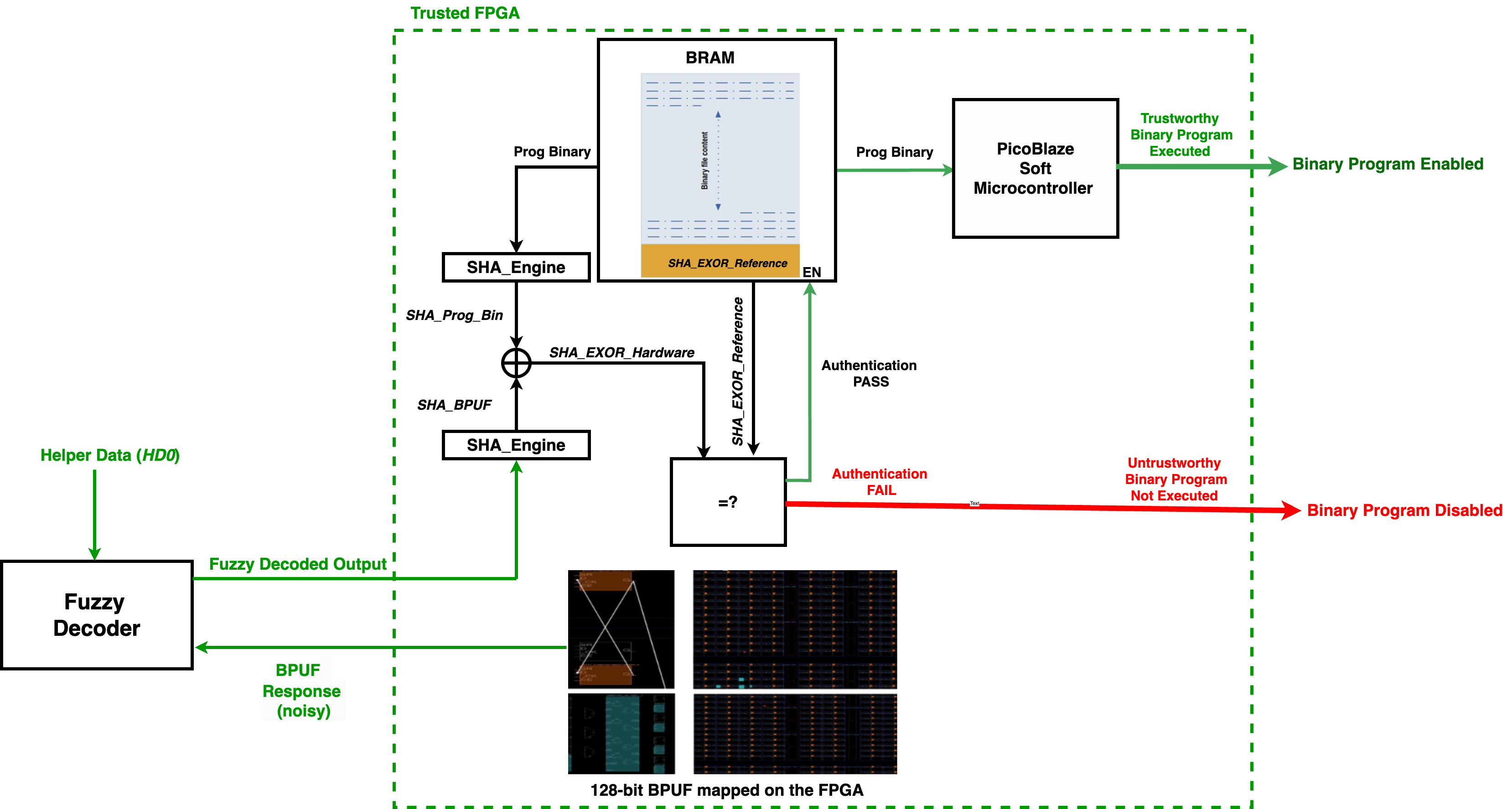}
\caption{BPUF-based program binary authentication details: success and failure
scenarios.}
\label{fig:BPUF_based_Prog_Bin_Authen}
\end{figure*}
\begin{figure*}[t]
\centering
\includegraphics[scale=0.23]{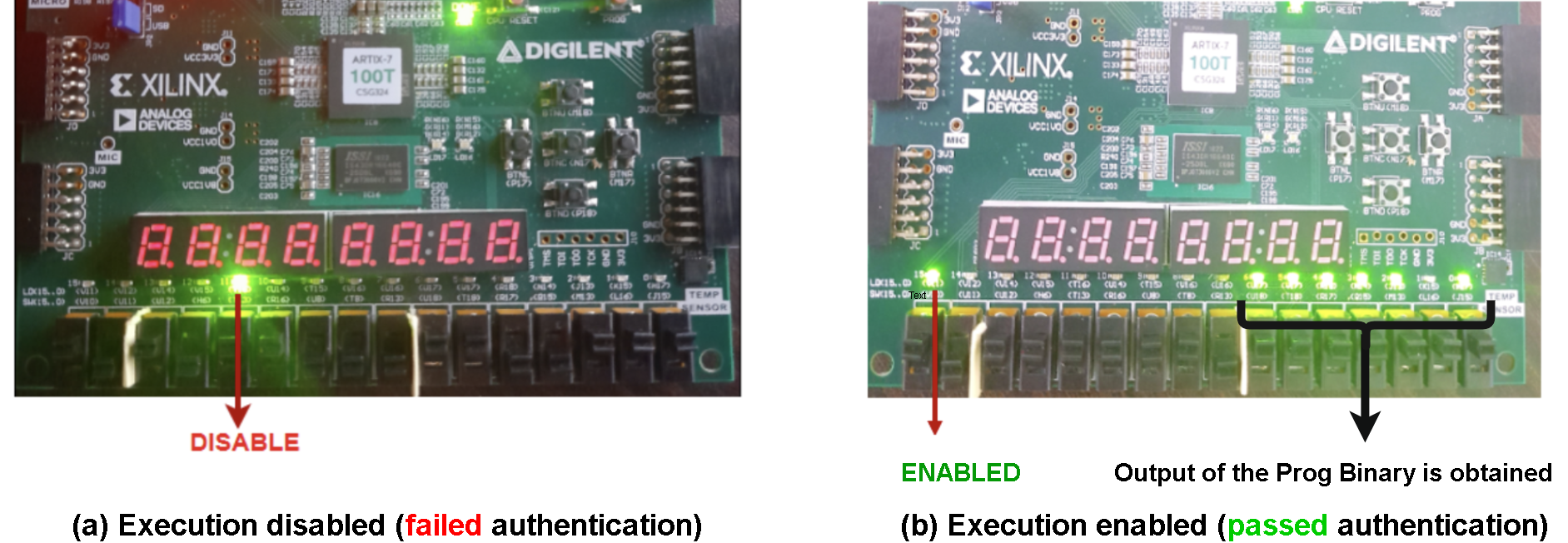}
\caption{Example response of the \emph{PUFBind} authentication system for 
(a) failed, and (b) successful platform and binary authentication.}
\label{fig:Compromised Program Binary Disabled from Execution}
\end{figure*}

Fig.~\ref{fig:Overall_Architecture_PUFBind} gives a 
detailed view of Step-2 of \emph{PUFBind}.
We describe the sub-steps in detail below.
\begin{enumerate} [label=\Alph*.]
\item \emph{Program Binary Generation:}
\label{item:program-binary-generation} 
A given assembly language program targeting 
the PicoBlaze microcontroller 
is assembled to generate a hexadecimal (.hex) file. 
Each PicoBlaze instruction is 18 bits long. The .hex file is 
edited by a Python script written by us to prepend 
sufficient zeroes before every instruction, so that the end 
of each instruction is aligned to the end of a 32-bit word,
to make it compatible with the word length of the FPGA's 
internal \emph{Block RAM} (BRAM) module (see 
Fig.~\ref{fig:Bin_File}).
Fig.~\ref{fig:Bin_File} shows an example modified 
program binary image to be loaded on the FPGA BRAM, 
where the BRAM has been configured to contain 32-bit 
words, and is of total size 4 kB (i.e. 1024 words). 
We emphasize that this is just an example -- our 
scheme and the corresponding hardware architecture 
are capable of handling program binaries of arbitrary 
size. 

\item \emph{Binding Authentic Program Binary to 
Trustworthy FPGA Hardware:}
\label{item:program binding}
A SHA-256 signature of the given trustworthy program binary 
is generated (by software means) and then EXOR-ed  with 
$SHA256\_K0$ (again by software means). This results 
in a ``golden signature'' $SHA\_EXOR\_Reference$ (see 
Fig.~\ref{fig:BPUF_based_Prog_Bin_Authen} 
and Fig.~\ref{fig:Bin_File}), that represents the
binding of the authentic program binary and its 
intended target trustworthy FPGA.
    
The binary file is appended with the necessary number of zeroes 
depending on the BRAM size, and then the 256-bit $SHA\_EXOR\_Reference$
value is then appended (piggybacked) to the binary file by 
another Python script, so that the total size of the binary file is
exactly equal to the size of the configured BRAM. 
The complete binary image file is then loaded on the FPGA BRAM.
\textbf{Note that this program binary modification
is a one-time operation for a given program binary and 
its given specific target FPGA.}

\item \emph{Mapping Soft Microcontroller on FPGA:}
\label{item:microcontroller_FPGA} 
The PicoBlaze soft microcontroller is instantiated on 
the FPGA.

\item \emph{Authentication of Program Binary:} 
Once the modified binary file is written to the BRAM of the 
FPGA, the authentication process ensues. A SHA-256 hardware 
engine mapped on the FPGA calculates the digest of the
``key'' generated from the
response of the FPGA-mapped BPUF, 
after it has 
been corrected through an error correction code. This BPUF digest 
calculated by the hardware is named \emph{SHA\_BPUF} 
(Fig.~\ref{fig:BPUF_based_Prog_Bin_Authen}). If 
the FPGA is authentic, the value of \emph{SHA\_BPUF} 
would be equal to \emph{SHA256\_K0}. Following this, 
the program binary stored in BRAM undergoes block-wise 
processing for digest calculation. The program binary 
can be of arbitrary size, and the SHA-256 engine 
processes it by collecting 512 bits of it (in our case it 
is 16 words of 32-bit each) in a 512-bit buffer.

After the final block of the program binary code has been 
processed (the portion appended with zeroes but excluding the
piggybacked 256-bit \emph{SHA\_EXOR\_Reference}), the 
output digest is referred to as \emph{$SHA\_Prog\_Bin$} 
(Fig.~\ref{fig:BPUF_based_Prog_Bin_Authen}). This sequential
reading and processing requires substantial synchronisation,
and relatively complex control circuitry for it to handle
arbitrary-sized program binaries.
Subsequently, the two digests \emph{SHA\_Prog\_Bin} and 
\emph{SHA\_BPUF} are EXOR-ed to generate a 256-bit value 
\emph{SHA\_EXOR\_Hardware}. The \emph{SHA\_EXOR\_Reference} 
value is then read from the BRAM and compared against
\emph{SHA\_EXOR\_Hardware}. If these two values match, 
authentication succeeds, reading of the program binary from 
the BRAM by the PicoBlaze microcontroller is enabled, and
normal execution of the program ensues. 
On the other hand, if authentication fails, access to the
program binary by the Picoblaze microcontroller remains
disabled, and an error condition is indicated by a single 
ON LED, as shown in
Fig.~\ref{fig:Compromised Program Binary Disabled from Execution}(a). 


\item \emph{Trustworthy Program Binary Execution:} If the program
binary has passed authentication, the PicoBlaze microcontroller 
mapped on the FPGA is allowed to access the BRAM to read the
program binary one instruction per clock cycle, and execute 
the instruction. Note that the first 14 zero bits of each word 
read from the BRAM are ignored, and only the last 18 bits
constitute the instruction. The result of normal execution 
may be indicated through multiple ON LEDs, as shown in
Fig.~\ref{fig:Compromised Program Binary Disabled from Execution}(b). 
\vspace{-8pt}
\subsection{Formal Proof of Security for \emph{PUFBind}}
We formally prove the security of the scheme by following 
the technique of ``proof by contradiction'', where we
assume the correctness of certain scenarios, and then 
show that our assumption leads to impossble conclusions. 
There are three possible cases that we must consider: \newline
\textbf{Case-1}: a modified program binary, with the 
\emph{SHA\_EXOR\_Reference} value corresponding to an 
authentic program binary piggybacked on it, passes
authetication on trustworthy FPGA. This would require
the following condition to be satisfied:
\vspace{-1pt}
\begin{multline}
SHA\_EXOR\_Reference =
        SHA\_Prog\_Bin'  \\
        \oplus SHA\_BPUF
\end{multline}
where $SHA\_Prog\_Bin'$ represents the SHA-256 digest
of the modified malicious program binary. This implies
the following endition must hold:
\begin{multline}
SHA\_Prog\_Bin \oplus SHA256\_K0 = \\ 
                 SHA\_Prog\_Bin'  \oplus SHA\_BPUF
\end{multline}
However, if the FPGA is trustworthy, \emph{SHA256\_K0} = \emph{SHA\_BPUF}.
Hence, the necessary condition is:
\begin{equation}
    SHA\_Prog\_Bin = SHA\_Prog\_Bin'
\end{equation}
which occurs with extremely low probability, because
SHA-256 is a collision-resistant hash function. Hence,
our assumption about the authentication step passing 
must be incorrect.

\textbf{Case-2:} an unmodified (trustworthy) 
program binary, with an unmodified value of 
\emph{SHA\_EXOR\_Reference} piggybacked on it, 
passes authetication on a untrustworthy FPGA
(on which it is not bound). 
This would require the following condition to 
be satisfied:
\begin{multline}
SHA\_Prog\_Bin \oplus SHA256\_K0 = \\ 
                 SHA\_Prog\_Bin  \oplus SHA\_BPUF'
\end{multline}
where \emph{SHA\_BPUF'} represents the signature computed
from the PUF on the untrustworthy FPGA. This
simplifies to:
\begin{equation}
SHA256\_K0 = SHA\_BPUF'
\end{equation}
which occurs with extremely low probability, 
since these are SHA-256 digests of the ``key'' 
values generated from the fuzzy extractor for 
the two PUFs located on
two different FPGAs. Hence, our assumption about
Case-2 must be incorrect.

\textbf{Case-3}: the program binary is unmodified
and the FPGA platform is truthworthy and bound to
the program binary, but the piggybacked signature
$SHA\_EXOR\_Reference$ has been modified to a
different value, (say) $SHA\_EXOR\_Reference'$. It 
is trivial to see that this case will fail
authentication, 
since $SHA\_EXOR\_Reference \neq SHA\_EXOR\_Reference'$.

Note that the other case where an untrusted program binary
is targeted to be executed on an untrusted FPGA is
irrelevant, because of our assumption in
Section~\ref{sec:salient_features}. 


\end{enumerate}

\section{Experimental Results}
\label{sec:results}
\subsection{Experimental Setup}
We implemented the proposed scheme on many Xilinx/AMD Artix-7 FPGA, 
hosted on a Digilent Nexys-A7 board. The hardware architecture 
(including the SHA-256 hardware engine and the controllers) 
of the proposed scheme was implemented in Verilog, and the 
UART transmitter (written in VHDL) was adopted from the KCPSM6 distribution
\cite{ref:AMD_picoblaze}. A 4 kB BRAM the FPGA was configured
to consist of 1024 words, each word being 32 bits long (see
Fig.~\ref{fig:Bin_File}). 
An open-source Python-based software 
Fuzzy Extractor~\cite{ref:fuzzy_extractor_software} was leveraged. 
Assembly language programs targeting the PicoBlaze microcontroller 
were assembled using the KCPSM6 toolchain, and Xilinx/AMD Vivado 
was used as the overall hardware design and debug environment. 
Custom Xilinx Design Constraint (.xdc) files were written to control
the placement of the 128-bit BPUF mapped on the FPGA, and to
map the on-board LEDs to circuit outputs.
All software steps, including processing of the BPUF collected response, 
were achieved through custom Python scripts.
A C program was used to automatically generate Verilog RTL
for the complex finite state machine (FSM) that controls 
the interaction between the SHA-256 hardware engine and 
the BRAM, since the number of states and state transitions 
of the FSM depends on the given program binary size. 
\subsection{Verification of Functionality} 
To validate the functionality of \emph{PUFBind}, three 
program binary images were construed: (a) an ``authentic'' 
program binary bound to the target FPGA, to make an array 
of on-board LEDs blink in a ring counter sequence; 
(b) a deliberately modified version of it, with a single 
arbitrary bit flip in some of the encoded instructions 
(``Case-1'' of the previous section), and (c) a modified 
binary image where a single arbitrary bit-flip was made 
in the piggybacked signature portion (``Case-3'' of the 
previous section).  Each of these three program
binaries were subjected to the proposed authentication flow. 
As expected, the first version passed authentication, while
the other two versions failed authentication. 
Authentication also failed when the program binary bound to 
a specific FPGA was attempted to be executed on a different 
FPGA (``Case-2'' of previous section). 
Fig.~\ref{fig:Compromised Program Binary Disabled from Execution} 
shows two scenarios -- for failed and passed authentication. 
Thus, the functionality of \emph{PUFBind} is verified.
\begin{table*}[t]
\centering
\caption{Hardware Cost of Proposed Architecture for Authentication (excluding PicoBlaze microcontroller)}
\scalebox{0.95}{%
\begin{tabular}{| c || c | c | c | c | c | c |}
\hline
\textbf{Name} & \textbf{Slice LUTs} & \textbf{Slice Registers} & \textbf{F7 Muxes} &  \textbf{Block RAM Tile} & \textbf{Bonded IOB}  & \textbf{BUFGCTRL} \\
{} & \textbf{(available: 63400)} & \textbf{(available: 126800)} & \textbf{(available: 31700)} & \textbf{(available: 135)} & \textbf{(available: 210)} & \textbf{(available: 32)} \\ 
\hline
\textbf{SHA\_Engine} & 1402 & 1057 & 64 & 0 & 0 & 0 \\ 
\hline
\textbf{128-bit BPUF}  & 0 & 256 & 0 & 0 & 0 & 0 \\ 
\hline
\textbf{Binary Program}  & 0 & 0 & 0 & 1& 0 & 0 \\ 
\hline
\textbf{Authenticator}  & 88 & 0 & 0 & 0 & 0 & 0 \\ 
\hline
\textbf{Custom Controller}  & 15 & 12 & 0 & 0 & 0 & 0 \\ 
\hline
\textbf{KCPSM6 Microcontroller} & 111 & 79 & 0 & 0 & 0& 0 \\ 
\hline
\textbf{UART}  & 55 & 155 & 8 & 0 & 0 & 0 \\ 
\hline \hline
\textbf{Total Utilization} (\%)  & 1671 (2.64\%) & 1559 (1.23\%) & 72 (0.2\%)& 1 (0.7\%)& 0 (0\%) & 0 (0\%) \\ 
\hline
\end{tabular}}
\label{table:ta}
\end{table*}
\subsection{Resource and Performance Costs}
Table~\ref{table:ta} shows the hardware footprint 
of the proposed security scheme (excluding the PicoBlaze microcontroller), 
which establishes the hardware overhead to be very small. There is no impact
on the critical path delay for the complete prototype system including 
the Picoblaze microcontroller, i.e. zero timing overhead.
For a BRAM with 32-bit words, it takes (512 / 32 = 16) 
clock cycles by the SHA-256 engine to process 512 bits of content 
from the BRAM at a time. Since the processes of filling-up the 
input 512-bit buffer and computation of the SHA-256 digest are concurrent,
for a 4 kB BRAM, it takes 16 * (64 blocks of 512 bits) = 1024 clock cycles
to compute the entire digest. One additional clock cycles is required
to compare the $SHA\_EXOR\_Reference$ and $SHA\_EXOR\_Hardware$. Hence,
for a 4 kB binary image, authentication takes 1025 clock cycles in total.
In the FPGA board we used, the system clock has a frequency of 100 MHz,
implying an one-time initial authentication latency of 102.5 microseconds.
Overall, the latency is more-or-less proportional to the binary image
size.

\subsection{Comparison of Overhead with Existing Schemes}
A qualitative comparison of \emph{PUFBind} with existing
solutions in terms of features, functionality, and applicability 
was presented in Section~\ref{sec:background}. 
A direct quantitative comparison between our proposed 
scheme and existing solutions is challenging since most
available solutions are for general-purpose processors,
use real-time encryption/decryption, and modifies the datapath. 
However, we note that the 128-bit BPUF hardware cost incurred is 
minuscule (only 2 flip-flops per bit), and the hardware resource
cost of the SHA-256 incurred by us compares favorably with the most 
hardware-efficient solutions reported in the comprehensive 
work~\cite{ref:Efficient_FPFA_SHA-256}.
Even if we had chosen to implement the  Fuzzy Extractor algorithm 
in hardware on FPGA in place of a hardware implementation, it would have
only cost us an additional $\sim237$ slices, and 4-kB 
additional BRAM~\cite{ref:puf_helper_data_algo}. 
We have provided a comparison between the 
hardware implementation costs of a reconfigurable TPM, 
and \emph{PUFBind} for an \emph{Artix-7} target platform
\cite{ref:reconfigurable_TPM,ref:TPM_hardware_Implementation} 
in Table~\ref{tab:COMPARISON}. Again, these results establish
the relative lightweightedness of our scheme. Note that it is difficult to
estimate the hardware costs of proprietary
technologies like \emph{TrustZone} and
\emph{Secure Enclave}.
\begin{table}[h!]
\centering
\caption{Comparison of Hardware Resource Utilization between 
TPM (FPGA Implementation) and \emph{PUFBind}}
\label{tab:COMPARISON}
\begin{tabular}{|p{3cm}|c|c|}
\hline
\textbf{TPM Components} & \textbf{LUTs Used} & \textbf{Slice Registers Used} \\
\hline
AES & 4487 (7.08\%) & 1162 (0.92\%) \\
\hline
SHA-1 & 1516  & 1587  \\
\hline
Modular Exponentiation & 1371  & 838  \\
\hline
PRNG & 348  & 205  \\
\hline
System Components & 612  & 699  \\
\hline
RSA & 22748 (35.88\%) & 24808 (19.56\%) \\
\hline
TPM Cost (Total) & \textbf{30842 (48.65\%)} & \textbf{29120 (22.97\%)} \\
\hline
\textbf{\emph{PUFBind} Cost (Total)} & \textbf{1671 (2.64\%)} & \textbf{1559 (1.23\%)} \\
\hline
\end{tabular}
\end{table}
\vspace{-8pt}
\section{Conclusions}
\label{sec:conclusions}
We have proposed and successfully demonstrated a practical, 
non-invasive hardware-software co-design scheme that securely 
binds a given program binary to a target FPGA using a PUF and 
cryptographic hash function. The proposed scheme eliminates 
the need for encryption and decryption of the program binary, 
thereby avoiding any performance penalties, and it does not 
require the storage of secret keys.
Our experimental prototype shows that the scheme introduces 
minimal hardware overhead and zero runtime performance overhead, 
aside from a one-time authentication latency, while offering 
high scalability and flexibility. In future work, we aim to 
extend this scheme to RISC-V processors.

\bibliographystyle{IEEEtran}
\bibliography{PUFBind}

\begin{thebibliography}{10}
\providecommand{\url}[1]{#1}
\csname url@samestyle\endcsname
\providecommand{\newblock}{\relax}
\providecommand{\bibinfo}[2]{#2}
\providecommand{\BIBentrySTDinterwordspacing}{\spaceskip=0pt\relax}
\providecommand{\BIBentryALTinterwordstretchfactor}{4}
\providecommand{\BIBentryALTinterwordspacing}{\spaceskip=\fontdimen2\font plus
\BIBentryALTinterwordstretchfactor\fontdimen3\font minus \fontdimen4\font\relax}
\providecommand{\BIBforeignlanguage}[2]{{%
\expandafter\ifx\csname l@#1\endcsname\relax
\typeout{** WARNING: IEEEtran.bst: No hyphenation pattern has been}%
\typeout{** loaded for the language `#1'. Using the pattern for}%
\typeout{** the default language instead.}%
\else
\language=\csname l@#1\endcsname
\fi
#2}}
\providecommand{\BIBdecl}{\relax}
\BIBdecl

\bibitem{ref:Renesas_RISC_V}
{RENESAS, Inc}, ``{RISC-V 32 \& 64-Bit MCUs and MPUs},'' \url{https://www.renesas.com/us/en/products/microcontrollers-microprocessors/risc-v}, 2024, {Accessed: September 2024}.

\bibitem{ref:puf_tutorial}
C.~Herder, M.-D. Yu, F.~Koushanfar, and S.~Devadas, ``{Physical Unclonable Functions and Applications: A Tutorial},'' \emph{Proceedings of the IEEE}, vol. 102, no.~8, pp. 1126--1141, 2014.

\bibitem{ref:security_in_FPGA_based_systems}
R.~Niranjana, ``{Security in FPGA-based Systems: Threats and Countermeasures},'' \url{https://www.logic-fruit.com/blog/fpga/security-in-fpga-based-systems/}, 2024, {Accessed: September 2024}.

\bibitem{ref:disassembler_1}
{hex-rays}, ``{IDA Pro: A powerful disassembler and a versatile debugger},'' \url{https://hex-rays.com/ida-pro/}, 2024, {Accessed: September 2024}.

\bibitem{ref:disassembler_2}
{HHD Software Ltd.}, ``{Free Hex Editor Neo},'' \url{https://hhdsoftware.com/free-hex-editor}, 2024, {Accessed: September 2024}.

\bibitem{ref:variable_length}
I.~Alshaer, B.~Colombier, C.~Deleuze, V.~Beroulle, and P.~Maistri, ``{Variable-Length Instruction Set: Feature or Bug?}'' in \emph{2022 25th Euromicro Conference on Digital System Design (DSD)}, 2022, pp. 464--471.

\bibitem{ref:fault_injection}
L.~Rivière, Z.~Najm, P.~Rauzy, J.-L. Danger, J.~Bringer, and L.~Sauvage, ``{High precision fault injections on the instruction cache of ARMv7-M architectures},'' in \emph{2015 IEEE International Symposium on Hardware Oriented Security and Trust (HOST)}, 2015, pp. 62--67.

\bibitem{ref:secure_firmware_update}
L.~Keleman, D.~Matić, M.~Popović, and I.~Kaštelan, ``{Secure firmware update in embedded systems},'' in \emph{2019 IEEE 9th International Conference on Consumer Electronics (ICCE-Berlin)}, 2019, pp. 16--19.

\bibitem{ref:stuxnet}
D.~Kushner, ``The real story of stuxnet,'' \emph{IEEE Spectrum}, vol.~50, no.~3, pp. 48--53, 2013.

\bibitem{ref:Solarwinds}
{Fortinet, Inc.}, ``{SolarWinds Cyber Attack: An Overview},'' \url{https://www.fortinet.com/resources/cyberglossary/solarwinds-cyber-attack}, 2024, {Accessed: September 2024}.

\bibitem{ref:hezbollah_explosions}
\BIBentryALTinterwordspacing
B.~News, ``Hezbollah pager devices linked to deadly explosions in september 2024,'' \url{https://www.bbc.com/news/articles/cz04m913m49o}, 2024, accessed: December 16, 2024. [Online]. Available: \url{https://www.bbc.com/news/articles/cz04m913m49o}
\BIBentrySTDinterwordspacing

\bibitem{ref:SPDM_formal_analysis}
\BIBentryALTinterwordspacing
C.~Cremers, A.~Dax, and A.~Naska, ``{Formal Analysis of {SPDM}: Security Protocol and Data Model version 1.2},'' in \emph{32nd USENIX Security Symposium (USENIX Security 23)}.\hskip 1em plus 0.5em minus 0.4em\relax Anaheim, CA: USENIX Association, 2023, pp. 6611--6628. [Online]. Available: \url{https://www.usenix.org/conference/usenixsecurity23/presentation/cremers-spdm}
\BIBentrySTDinterwordspacing

\bibitem{ref:TPM_book}
W.~Arthur and D.~Challener, \emph{{A Practical Guide to TPM 2.0: Using the Trusted Platform Module in the New Age of Security}}, 1st~ed.\hskip 1em plus 0.5em minus 0.4em\relax USA: Apress, 2015.

\bibitem{ref:TPM_intel}
{Intel, Inc.}, ``{Trusted Platform Module (TPM) Overview},'' \url{https://www.intel.com/content/www/us/en/business/enterprise-computers/resources/trusted-platform-module.html}, 2024, {Accessed: September 2024}.

\bibitem{ref:secure_enclave_apple}
{Apple, Inc.}, ``{Apple Platform SEcurity: Secure Enclave},'' \url{https://support.apple.com/en-in/guide/security/sec59b0b31ff/web}, 2024, {Accessed: September 2024}.

\bibitem{ref:trustzone_arm}
{Arm}, ``{ARM TrustZone technology},'' \url{https://developer.arm.com/documentation/100690/0200/ARM-TrustZone-technology}, 2024, {Accessed: September 2024}.

\bibitem{ref:ti_firmware}
{Texas Instruments Inc.}, ``{System Firmware Authentication and Decryption Requests},'' \url{https://software-dl.ti.com/tisci/esd/latest/6_topic_user_guides/authentication.html}, 2016, {Accessed: September 2024}.

\bibitem{ref:AEGIS_processor}
G.~E. Suh, D.~Clarke, B.~Gassend, M.~van Dijk, and S.~Devadas, ``{AEGIS: architecture for tamper-evident and tamper-resistant processing},'' in \emph{Proceedings of the 17th Annual International Conference on Supercomputing}, ser. ICS '03, 2003, p. 160–171.

\bibitem{ref:intel_tdx}
{Intel Corporation}, ``{Intel Trust Domain Extensions (Intel TDX): Isolation, confidentiality, and integrity at the virtual machine (VM) level},'' \url{https://www.intel.com/content/www/us/en/developer/tools/trust-domain-extensions/overview.html}, 2022, {Accessed: September 2024}.

\bibitem{ref:puf_Gassend2002}
B.~Gassend, D.~Clarke, M.~van Dijk, and S.~Devadas, ``{Silicon physical random functions},'' in \emph{{Proc. of ACM Conference on Computer and Communications Security}}.\hskip 1em plus 0.5em minus 0.4em\relax New York, NY, USA: ACM Press, 2002, pp. 148--160.

\bibitem{ref:puf_synopsys}
{Synopsys}, ``{Enabling Authentication with PUF-based Security},'' \url{https://www.synopsys.com/designware-ip/security-ip/cryptography-ip/puf.html}, 2024, {Accessed: September 2024}.

\bibitem{ref:neopuf}
{eMemory Technology Inc.}, ``{NeoPUF},'' \url{https://www.ememory.com.tw/en-US/Products/Product?guid=19081314113656}, 2019, {Accessed: September 2024}.

\bibitem{ref:pufcc}
{PUFsecurity}, ``{PUFcc: A PUF-based Crypto Coprocessor},'' \url{https://www.pufsecurity.com/products/pufcc/}, 2019, {Accessed: September 2024}.

\bibitem{ref:synopsys_puf}
{Synopsys, Inc.}, ``{Enabling Authentication with PUF-based Security},'' \url{https://www.synopsys.com/designware-ip/security-ip/cryptography-ip/puf.html}, 2024, {Accessed: September 2024}.

\bibitem{ref:synopsys_puf_vault}
------, ``{Synopsys PUF IP for Key Vault},'' \url{https://www.synopsys.com/designware-ip/security-ip/cryptography-ip/puf/key-vault.html}, 2024, {Accessed: September 2024}.

\bibitem{ref:synopsys_firm_IP}
------, ``{Firmware IP Protection: Protection against Reverse-Engineering, Counterfeiting/Cloning and Overbuilding},'' \url{https://www.synopsys.com/designware-ip/security-ip/cryptography-ip/puf/firmware-ip-protection.html}, 2024, {Accessed: September 2024}.

\bibitem{ref:synopsys_puf_software}
{Synopsys}, ``{Synopsys PUF Base/Premium - Software},'' {\url{https://www.synopsys.com/dw/ipdir.php?ds=security-puf-ip-software}}, 2024, {Accessed: September 2024}.

\bibitem{ref:puf_IoT_protocol}
S.-W. Lee, M.~Safkhani, Q.~H. Le, O.~H. Ahmed, M.~Hosseinzadeh, A.~M. Rahmani, and N.~Bagheri, ``{Designing secure PUF-based authentication protocols for constrained environments},'' \emph{Scientific Reports}, vol.~13, no.~1, 2023.

\bibitem{ref:butterfly_PUF}
S.~Kumar, J.~Guajardo, R.~Maes, G.-J. Schrijen, and P.~Tuyls, ``{Extended abstract: The butterfly PUF protecting IP on every FPGA},'' in \emph{{Proc. of IEEE International Workshop on Hardware-Oriented Security and Trust(HOST)}}, June 2008, pp. 67--70.

\bibitem{ref:anderson_puf}
J.~H. Anderson, ``{A PUF design for secure FPGA-based embedded systems},'' in \emph{2010 15th Asia and South Pacific Design Automation Conference (ASP-DAC)}, 2010, pp. 1--6.

\bibitem{ref:sram_puf_fpga}
A.~Wild and T.~Güneysu, ``{Enabling SRAM-PUFs on Xilinx FPGAs},'' in \emph{2014 24th International Conference on Field Programmable Logic and Applications (FPL)}, 2014, pp. 1--4.

\bibitem{ref:neopuf_on_fpga}
{eMemory Technology Inc.}, ``{Achronix Adopts eMemory IP for FPGA Hardware Root of Trust},'' \url{https://www.ememory.com.tw/en-US/News/News?guid=21042815130660}, 2021, {Accessed: September 2024}.

\bibitem{ref:puf_helper_data_algo}
R.~Maes, P.~Tuyls, and I.~Verbauwhede, ``Low-overhead implementation of a soft decision helper data algorithm for sram pufs,'' in \emph{Cryptographic Hardware and Embedded Systems - CHES 2009}, C.~Clavier and K.~Gaj, Eds.\hskip 1em plus 0.5em minus 0.4em\relax Berlin, Heidelberg: Springer Berlin Heidelberg, 2009, pp. 332--347.

\bibitem{ref:fuzzy_extractor_theory}
R.~Canetti, B.~Fuller, O.~Paneth, L.~Reyzin, and A.~Smith, ``{Reusable Fuzzy Extractors for Low-Entropy Distributions},'' \emph{Journal of Cryptology}, vol.~34, no.~1, Nov 2020.

\bibitem{ref:synopsys_butterfly_puf}
{Synopsys, Inc.}, ``{Synopsys PUF FPGA-X Base/Premium},'' \url{https://www.synopsys.com/dw/ipdir.php?ds=security-puf-fpga-x}, 2024, {Accessed: September 2024}.

\bibitem{ref:fuzzy_extractor_software}
{Yageman, Carter}, ``{A Python implementation of Fuzzy Extractor},'' \url{https://github.com/carter-yagemann/python-fuzzy-extractor}, {2018}, {Accessed: September 2024}.

\bibitem{ref:AMD_picoblaze}
I.~Advanced Micro~Devices, ``{PicoBlaze 8-bit Microcontroller},'' \url{https://www.xilinx.com/products/intellectual-property/picoblaze.html#overview}, 2014, {Accessed: September 2024}.

\bibitem{ref:Efficient_FPFA_SHA-256}
H.~E. Michail \emph{et~al.}, ``{On the exploitation of a high-throughput SHA-256 FPGA design for HMAC},'' \emph{ACM Trans. Reconfigurable Technol. Syst.}, vol.~5, no.~1, mar 2012.

\bibitem{ref:reconfigurable_TPM}
M.~D. James, ``{A Reconfigurable Trusted Platform Module},'' Master's thesis, USA, 2017, aAI28104420.

\bibitem{ref:TPM_hardware_Implementation}
Z.~Liu, Y.~Wang, and T.~Liu, ``{Research on the Implementation of Trusted Platform Module based on Reconfigurable Computing},'' in \emph{2010 International Conference on Anti-Counterfeiting, Security and Identification}, 2010, pp. 85--87.

\end{thebibliography}

\end{document}